# A TABULATION AND CRITICAL ANALYSIS OF THE WAVELENGTH-DEPENDENT DIELECTRIC IMAGE COEFFICIENT FOR THE INTERACTION EXERTED BY A SURFACE ONTO A NEIGHBOURING EXCITED ATOM

Solomon Saltiel [(1,2)], Daniel Bloch[(1)*] and Martial Ducloy[(1)]

[(1)] *Laboratoire de Physique des Lasers, UMR7538 du CNRS et de l'Université Paris13. 99, Av. J.B. Clément, F-93430 Villetaneuse, France*

[(2)] *Physics Department, Sofia University, 5, J. Bourchier Blvd., 1164 Sofia, Bulgaria*

* *e-mail: bloch@lpl.univ-paris13.fr*

**Abstract**

*The near-field interaction of an atom with a dielectric surface is inversely proportional to the cube to the distance to the surface, and its coupling strength depends on a dielectric image coefficient. This coefficient, simply given in a pure electrostatic approach by* $(\varepsilon-1)/(\varepsilon+1)$ *with* $\varepsilon$ *the permittivity, is specific to the frequency of each of the various relevant atomic transition : it depends in a complex manner from the bulk material properties, and can exhibit resonances connected to the surface polariton modes. We list here the surface resonances for about a hundred of optical windows whose bulk properties are currently tabulated. The study concentrates on the infrared domain because it is the most relevant for atom-surface interaction. Aside from this tabulation, we discuss simple hints to estimate the position of surface resonances, and how uncertainties in the bulk data for the material dramatically affect the predictions for the image coefficient. We also evaluate the contribution of UV resonances of the material to the non resonant part of the image coefficient.*

## 1. Introduction

Atomic Physics and the related high-resolution sensitive spectroscopy techniques allow for the probing of long-range atom-surface interaction [1] with a high accuracy. Recently, it has been experimentally demonstrated that the universal van der Waals (vW) attraction between an atom and a neighbouring surface, that spans in $z^{-3}$ with z the atom-surface distance, could be turned into a repulsion [2,3] through a resonant coupling between virtual atomic transitions and resonances of the surface. It was also shown [4] that in a related process, an excited atom can undergo a remote quenching to a lower energy state analogous to a Förster-type energy transfer here applied to the surface mode. The long-range coupling to the surface can indeed open an energy-transfer channel, that would remain otherwise nearly prohibited for spontaneous emission in the vacuum. More generally, the development of various techniques confining cold atoms close to surfaces and the attempts to selectively deposit atoms or thin layers for nanofabrication purposes, induce a growing need for the control and engineering of the atom-surface interaction.

It is the purpose of this paper to provide in a simple manner, and for a large set of materials, the surface-related parameters determining the atom-surface interaction. Because the atom-surface interaction can be expanded over the various atomic transitions to coupled levels, the specific properties of the considered dense material can be determined by a simple "*image coefficient*" (relative to an ideal reflecting surface), defined for each relevant atomic coupling. As recalled below, these coefficients are in the principle deduced from the spectral knowledge of the bulk permittivity of the material $\varepsilon(\omega)$, through a complex (planar) surface response function S, that simply turns to be $S(\varepsilon) = (\varepsilon - 1) / (\varepsilon + 1)$ for a non-dispersive material.

The paper is presented in the following way. In section 2, we briefly recall the essential results for the physics of the atom-surface interaction in the near-field regime, in order to provide in an intelligible manner the reflection coefficients applicable for a virtual transition in *absorption*, as well as for a virtual *emission*, and the dielectric coefficient relevant for a *real* energy transfer. Emphasis is on these atomic *emission* processes –occurring only for *excited* atoms-, as they are susceptible to couple resonantly with the surface mode resonances naturally appearing in the surface response function S [5]. Section 3 is mostly devoted to a listing of the surface resonances obtained for a large list of optical materials, essentially those whose bulk values are known from the Palik Handbook [6] tabulation, or for which a fitting expression for $\varepsilon(\omega)$ is published in the literature. We concentrate on resonances in the IR domain, and hence on dielectric and semi-conductor materials, because the IR contributions usually provide the dominant vW surface interaction, and because "optical" materials most often exhibit a transparency window in the visible range (*i.e.* no resonance should appear in the visible). Although the presented results are derived from a numerical evaluation, we also discuss a simple method to approximately locate the resonance. Section 4 discusses the issue of accuracy of the predictions for a resonant behaviour, showing that apparently minor discrepancies between published data for the bulk material may lead to dramatically differing predictions for the surface behaviour. This is illustrated with the examples of AlSb, InSb and YAG, and then discussed on a more general basis : in particular, it is shown that the original data -notably reflectivity studies- from which the bulk permittivity is usually extracted, can be more relevant than the use of tabulated or spectrally modelled values of permittivity. Aside from the resonant behaviour, an accurate determination of the nonresonant contributions can also be needed, notably because the effective atomic behaviour usually results from a summing of various contributions, most of them non resonant. This implies that the specific "exotic" behaviour (*e.g.* repulsion) induced by a resonant term can be strongly corrected by the additional non resonant terms : section 5 concentrates on the smoothly frequency-varying non resonant contribution from tabulated bulk values. It is in particular shown that the UV resonances, although effective only through their far wings response, are far from being entirely negligible.

## 2. Atom interaction with a dielectric medium in the range of near-field electrostatic approximation

### 2.A Energy shift and virtual transitions

In the vicinity of a perfect reflector, an atom in the level $|i\rangle$ undergoes a dipole-dipole interaction expressed as :

$$V_i(z) = -\frac{1}{16z^3}\sum_j\left(\left|\mu^{ij}\right|^2 + \left|\mu_z^{ij}\right|^2\right) \qquad (1)$$

with $\mu^{ij}$ the dipole moments related to the virtual transition $|i\rangle \rightarrow |j\rangle$ . In eq (1), we assume that the retardation effects are negligible in order to ensure the electrostatic approximation, *i.e.* $z \ll \lambda_{ij}$, with $\lambda_{ij}$ the wavelength of the $|i\rangle \rightarrow |j\rangle$ transition; also, we typically assume $z \geq$ 1 nm, in order to be insensitive to the structural details of the surface. Such a description, with its $z^{-3}$ spatial dependence, is known to characterize the non-retarded atom-surface vW interaction, often described as the dipole coupling between a (fluctuating) atom dipole and its (instantaneously correlated) image induced in the reflecting surface. Note that through the summing over the dipole couplings appearing in eq.(1), the influence of IR transitions between atomic levels of neighbouring energy is strongly enhanced relatively to their relative weight in a spontaneous emission process (see [1]). This is why in the following, our focusing will be on the IR resonances of materials.

If the neighboring surface is not a perfect reflector, but a dielectric medium, the energy shift has to be modified in the following way:

$$V_i(z) = -\frac{1}{16z^3}\sum_n\left(\left|\mu^{ij}\right|^2 + \left|\mu_z^{ij}\right|^2\right)\, r(\omega_{ij}) \qquad (2)$$

with $r(\omega_{ij})$ a "dielectric image coefficient" affecting the virtual transition $|i\rangle \to |j\rangle$. If the dispersion of the dielectric medium could be neglected (i.e. the dielectric permittivity $\varepsilon$ is constant over the whole spectrum), this dielectric coefficient would be frequency-independent and simply given by the electrostatic image coefficient $r = (\varepsilon - 1) / (\varepsilon + 1)$. More correctly, when the dispersive features of the dielectric coefficient are taken into account [7], one finds for a virtual absorption (*i.e.* $\omega_{ij} > 0$) :

$$r_a(\omega_{ij}) = \frac{2}{\pi}\int_0^{\infty}\frac{\omega_{ij}}{\omega_{ij}^2 + u^2}\,\frac{\varepsilon(iu) - 1}{\varepsilon(iu) + 1}\,du \qquad (3)$$

and for a virtual emission *(i.e.* $\omega_{ij} < 0$) :

$$r_e(\omega_{ij}) = \frac{2}{\pi}\int_0^{\infty}\frac{\omega_{ij}}{\omega_{ij}^2 + u^2}\,\frac{\varepsilon(iu) - 1}{\varepsilon(iu) + 1}\,du + 2\Re e\left[\frac{\varepsilon(|\omega_{ij}|) - 1}{\varepsilon(|\omega_{ij}|) + 1}\right] \qquad (4)$$

Eq. (4) can be also written as :

$$r_e(\omega_{ij}) = -r_a(|\omega_{ij}|) + 2\Re e S(|\omega_{ij}|) \qquad (5)$$

where we have introduced in (5) the surface response function $S(\omega) = [\varepsilon(\omega) - 1]/ [\varepsilon(\omega) + 1]$.

From eqs. (3-5) one notes that, once the dispersion of the dielectric permittivity is taken into account, the knowledge of the permittivity on the whole spectrum is required. However, in eq. (3), *i.e.* for the case of a virtual *absorption* ($\omega_{ij} > 0$), causality and the Kramers-Krönig relationship impose the boundaries $0 < r(\omega) < 1$ along with a monotone behaviour for $r(\omega)$ as a function of $\omega$. Hence, one understands that the accuracy on $r(\omega)$ depends only smoothly upon the uncertainties in the determination of $\varepsilon(\omega)$. Conversely, for a virtual *emission* of the atom ($\omega_{ij} < 0$, eqs. (4-5)), there appears a second term in the dielectric coefficient that is susceptible to evolve arbitrarily: its amplitude can possibly exceed unity, its sign can be positive or negative. These features have been analyzed [7,8] as originating in a resonance between a virtual absorption into a surface-plasmon [7] or a surface-polariton mode [8], and the atom emission. They are strongly dependent upon the spectral features of the dielectric medium.

2.B Surface-modified decay rate

In addition to the energy-shift induced by the vicinity with the surface, which even affects an atom in its ground-state, the decay rate of an excited atom, and the relative efficiency of the various de-excitation channels, can depend sharply on the vicinity with a surface. For our discussion, centred on the resonant effects, we do not consider the finite increase of the decay rate in the presence of a transparent dielectric surface, related with an enhanced spontaneous emission through the near-field evanescent-wave coupling between the emitting atom and the surface [1,9]. Rather, we consider the case when the bulk material is absorbing at the frequency associated to an IR transition between the excited atomic level and a neighbouring lower energy level. This decay channel -usually in the mid-IR range and hence often very weak for an atom in the vacuum- undergoes a strong $z^{-3}$ magnification in the vicinity with the surface, through a dissipative analogous of the resonant enhancement of the van der Waals interaction [1,4,7]. The atomic decay rate $\gamma_{ij}$ for the $|i> \rightarrow |j>$ process varies as :

$$\gamma_{ij}(z) = \gamma_{ij}(\infty)\left[1+(2\pi z/\lambda_{ij})^{-3}\Im m\left(\frac{\varepsilon(|\omega_{ij}|)-1}{\varepsilon(|\omega_{ij}|)+1}\right)\right] \qquad (6)$$

The notable result of eq.(6) is the appearance of the factor $\Im m\left[\left(\varepsilon(|\omega_{ij}|)-1\right)/\left(\varepsilon(|\omega_{ij}|)+1\right)\right]$ = $\Im m[S(|\omega_{ij}|)]$, which is the dissipative counterpart of the resonant term $\Re e S(|\omega_{ij}|)$ involved in eqs. (4-5). This $\Im m[S(|\omega_{ij}|)]$ factor governs the distance at which the surface-induced decay channel becomes predominant relatively to standard spontaneous emission.

## 3. Surface resonances of materials

As discussed in section 2, the most "exotic" behaviors induced by a resonant coupling between the atomic excitation and the surface polariton mode, are characterized by the complex surface response $S(\omega_{ij})$ as defined following eq.(5). Conversely, the non resonant contribution $r_a(|\omega_{ij}|)$ provides a contribution varying only smoothly with the energy of the atomic transition. These terms remain however important in the final summing of all virtual contributions, and cannot be ignored in the final assessment of the surface interaction.

As it is well-known, and will be further exemplified in section 4, $\Re e[S(\omega)]$ is essentially dispersion-like, and $\Im m[S(\omega)]$ absorption-like. A simplified modeling of the permittivity $\varepsilon(\omega)$ - notably those extrapolated from a dilute medium approach- would fully justify this point. Such a view is only approximate because resonances in dense media are much broader than current atomic resonances, and because the overlap of several neighbouring resonances most often precludes a perfect (anti-)symmetry. Nevertheless, a bunch of useful information can be described with the position, width, and amplitude of these resonances.

Figure 1 and Table 1 constitute the core of the paper, and characterize the surface resonances for numerous optical materials. The values of the bulk permittivity $\varepsilon(\omega)$ are mostly taken from the compiled values provided in the Palik handbook [6] or from fitting expressions for $\varepsilon(\omega)$ ; for birefringent materials, the permittivity is obtained by taking $(\varepsilon_{//}\varepsilon_{\perp})^{1/2}$ [10], the value that applies for a symmetry axis oriented towards the normal to the surface, making the cylindrical symmetry not broken in spite of the birefringence. As already discussed at length for the case of sapphire in [8], the surface resonances actually occur for radically differing frequencies than those of the bulk material. For the clarity of presentation, in figure 1, we have defined the position of the resonance(s) of a given material as the frequency associated to the peak value of the nearly absorption-like $\Im m[S(\omega)]$ ; for the amplitude, we characterize $\Im m[S(\omega)]$ by its peak value (one has $\Im m[S(\omega)] \geq 0$, and $\Im m[S(\omega)] = 0$ in the transparency window), and the nearly dispersion-like $\Re e[S(\omega)]$ by its extreme values. Note that these extreme values would be opposite and simply related to the $\Im m[S(\omega)]$ amplitude in the case of an ideal narrow and well-isolated resonance. In addition and to further characterize a resonance with an indication of its width, Table 1 provides the frequency positions of the extreme peaks of $\Re e[S(\omega)]$. Aside from these essential features, the general behavior of these resonances, including their far extended wings, can be calculated by directly applying the tabulated values of the complex index $n+i\kappa$ to evaluate $S(\omega)$ (through $\varepsilon = (n+i\kappa)^2$).

The information provided in table 1 and figure1, should make easy the selection of the right material if a resonance with a specific atomic excitation is needed. Oppositely, it also allows one to predict when the effect of a narrow resonance can be ruled out. In all cases, one has to keep in mind that the dispersive resonance for $\Re e[S(\omega)]$ implies slowly decaying tails, so that approximate coincidences, leading to resonant behaviours, are relatively easy to find. Note also that the resonant nature of the atom-surface van der Waals is truly dominant only when $|\Re e[S(\omega)]|$ is at least comparable with unity -or with $r_a(\omega)$ -. Conversely, at the smaller distance, even a relatively small value for $\Im m[S(\omega)]$ induces large changes of the lifetime and branching ratios : this is because there is no equivalent of a "non resonant" change for this dissipative effect.

Aside from these numerical evaluations, it is possible to assess an approximate location of the $S(\omega)$ resonances. Before, it is worth nothing that in the theory, the resonance is obtained for a pole of $[\varepsilon(\omega)+1]^{-1}$, but that this pole is at a complex frequency. Surface resonances are usually so broad that the complex pole frequency is not very useful for a practical location of surface resonances, notably the tiny ones.

As can be seen from Table 1, the "centre" of the resonance, as defined through the peak frequency of $\Im m[S(\omega)]$, is very close to the centre of the anomalous dispersion (for $\Re e[S(\omega)]$), and in most cases (for pronounced resonances) close to the zero value of $\Re e[S(\omega)]$ (see figure 1). With the complex permittivity $\varepsilon$ provided through the complex index $(n+i\kappa)$, one gets :

$$S(\omega)=\frac{\varepsilon(\omega)-1}{\varepsilon(\omega)+1}=1-\frac{2}{(n^2-\kappa^2+1)+2in\kappa}=\frac{(n^2+\kappa^2)^2-1+4in\kappa}{(n^2-\kappa^2+1)^2+4n^2\kappa^2} \quad (7)$$

so that the "resonance" (when defined by $\Re e[S(\omega)]=0$) occurs for :

$$n^2+\kappa^2 = 1 \quad (8)$$

With this relation, one easily shows that $\Im m[S(\omega_{res})]= \kappa/n$, and surface resonances will appear only if $n$ is small enough, and thus $\kappa$ ( $=\sqrt{1-n^2}$ ) close to unity. If $n<<1$, the resonance amplitude is on the order of $1/n$. It is also worth noting that the so-defined resonance condition can be read as $|\varepsilon|=1$, a condition that is satisfied by the pole condition (for complex frequency) $\varepsilon = -1$.

The interest for such a simple estimate is twofold : on the one hand, it provides, in a very elementary manner, a way to locate and characterize a surface resonance from the knowledge of optical values characterizing the bulk material; moreover, this estimate does not depend of a specific modeling of the bulk resonance. On the other hand, it shows that these surface resonances always occur in a frequency region where the optical material is strongly absorbing (typically on half a reduced wavelength), so that the material is no longer an optical "window", implying specific difficulties in the evaluation of its optical constants. In the next section, we discuss the issue of the uncertainties in the tabulated data, with respect to the fact that the exact features of the surface response $S(\omega)$ - and notably the sign of $\Re e[S(\omega)]$, upon which is based the prediction of a vW attraction or repulsion- are strongly dependent on the accuracy of the determination of n and $\kappa$.

## 4. Selecting bulk data to evaluate the surface resonance

It is naturally not an uncommon situation that measurements performed by various authors for the same material lead to accidental differences in the tabulated optical constants. The use of different samples, or differing experimental conditions, such as the temperature of the sample, may unsurprisingly lead to some discrepancies. More fundamentally, the spectral determination of a pair of optical constants $(n,\kappa)$ that are experimentally intricate, usually demands an amount of extrapolation. When the evaluation relies on the Kramers-Kronig relationship, the knowledge of the whole spectrum is even requested. However, when the goal of these optical analyses on the bulk material is to determine the volume resonances of a material, the final discrepancies usually appear to be relatively minor and insensitive to the absolute calibration of the optical measurements. Conversely, these marginal uncertainties lead to dramatic changes for *surface* resonances.

We illustrate below such situations. As a first example, we consider the case of AlSb, that features a single resonance in the far IR, and for which two sets of data for $(n,\kappa)$ are provided in [6], based upon two different experimental studies [14,15]. In figure 2a, the comparison of the plotted values for n and $\kappa$ according to the two different sets of data exhibits notable differences in some values, but no major discrepancies in the position of the peaks for these bulk parameters. However, the frequency where $\kappa \sim 1$ is strongly dependent on the choice of data. This explains that, as shown by fig.2b, the location of the predicted surface resonances is critically dependent on the considered set of $(n,\kappa)$ values. Conversely, the resonant behaviour of $S(\omega)$ in the wings of the surface resonance appears independent of the quality of the bulk data. Also, an analytical modeling of the bulk resonances (*e.g.* classical theory of dispersion), involving a limited number of parameters can be considered [14,15]: it usually leads to slightly modified values of the $(n, \kappa)$ set and to slightly sharper surface resonances, but does not essentialy alter the position of the surface resonances as deduced from an extrapolation of the tabulated values in [6]. A second illustration is provided by InSb, with two bulk resonances in the far IR : although the discrepancies occurring between the two sets of data (Fig.3a) are comparable for both resonances, one notices (fig. 3b) that one of the surface resonances (the one with the lower energy around 70 $cm^{-1}$, effectively measured by a Fourier transform method in [40], otherwise only extrapolated from IR data in [41]) is much more sensitive than the other one (around 190 $cm^{-1}$) to the choice of the set of bulk parameters. As an additional example, in the less remote IR range, YAG is a genuine dielectric (non semi-conductor) medium of a great practical importance (including for our own experiments with the vW interaction, see [3]) : it exhibits multiple bulk resonances, partly shown in fig. 4a. As for InSb, some of the surface resonances (fig. 4b) are extremely sensitive to the exact assessment of the bulk resonances. In addition to these simple illustrative examples, similar remarks could be derived from the differing sets of (n,κ) values found for example for GaAs, or for $BaF_2$, although some critical considerations may help to choose among the data proposed in the literature (see table 1).

As already mentioned, the set of (n,κ) value is usually not directly measured, and requires a disentanglement to be obtained. Among the current techniques to get these (n,κ) values, the measurement of reflectivity close to the normal incidence appears to be particularly relevant for these issues of surface resonances. It is possible to reconstruct the reflectivity from the (n,κ) data, given either by discrete tabulated values, or by an analytical modelling. As shown in fig 2c, 3c, 4c, a correlation appears between the sensitivity of the reflection spectrum to the considered set of data, and the predictions for the surface resonance. In most cases, the strongest disagreement between various sets of data is not for the position of the peaks of reflectivity, but rather occurs in the sharp wings of the reflectivity spectrum : there can be some discrepancies in the absolute values of reflectivities around the peaks, or in the typical "width" of the reflectivity resonance, but the most radical variations appear in the reflectivity values around these wings when comparing various sets of data. This connection between reflectivity and the surface response, can be understood from the Fresnel formulae for normal incidence. The reflectivity (in intensity) R(ω) being given by:

$$R(\omega)=\left|\frac{n+i\kappa-1}{n+i\kappa+1}\right|^2=\frac{(n-1)^2+\kappa^2}{(n+1)^2+\kappa^2}=1-\frac{4n}{(n+1)^2+\kappa^2} \qquad (9)$$

one sees that $R(\omega) \sim 1$ in the regions of strong bulk absorption (characterized by $\kappa >> 1$), while close to a surface resonance- eq.(8), one has rather $R(\omega) \sim (1-n)/(1+n)$. If sharp surface resonances are characterized by $\kappa \sim 1$, and $n << 1$, however, most of the surface resonances, when not an extremely sharp one, rather occur for $\kappa \leq 1$ and an arbitrary value of n ($n \leq 1$). In some cases, the experimental data directly measure the reflectivity, with uncertainties mostly originating from the absolute reflectivity calibration (e.g. for non evacuated systems, at wavelengths known for air absorption), or possibly from the wavelength selection system (especially for older apparatus), or from the imperfections of the surface state, responsible for a possible light scattering (although scattering losses are expected to be small in the IR range). These remarks show that when the literature is not precise enough to provide a reliable value of the resonant behaviour at a given wavelength, it should be sufficient to measure around the wavelength of interest the reflectivity of the window, in conditions (*e.g.* temperature) similar as close as possible as those used for the planned experiments. In this spirit, we had performed reflectivity measurements of two YAG windows on vapour cells currently used for our studies (fig. 4). They tend to establish that the data of ref [37] (used for our predictions in [3] for the ~ 820 $cm^{-1}$ resonance), is most probably irrelevant, at least for the YAG samples that we use.

**5. The non resonant contribution $r_a(\omega)$ and the influence of the UV absorption**

As recalled in section 3, the non resonant contribution $r_a(\omega)$ exhibits a smooth monotone decrease with $\omega$. Its intrinsic integration of fluctuation properties over the whole spectrum makes it remarkably insensitive to the uncertainties affecting the bulk properties. However, the evaluation of the precise behaviour of an atom - in a given state- in front of a surface, with its summing over numerous coupling transitions, may demand some accuracy in the evaluation of the $r_a(\omega)$ values. We discuss here some of the possible approaches for the evaluation of $r_a(\omega)$.

Because of the relative insensitivity of $r_a(\omega)$ to the details of the bulk permittivity, and because of the imaginary frequency appearing in eq.(3), it is very convenient to use, when available, an analytical expression for $\varepsilon(\omega)$, enabling an easy extension and calculation in the complex plane. However, in most cases (one of the few exceptions is for sapphire, see [13]), these analytical expressions are limited to the band of IR absorption band, and are irrelevant inside the transparency window, or in the UV absorption band. In the absence of an experimentally determined analytical expression spanning over the whole spectrum, $r_a(\omega)$ is numerically evaluated from its real-valued equivalent expression [8]:

$$r_a(\omega_0) = \frac{2}{\pi} P\int_0^{\infty} \Im m[S(\omega)] \frac{\omega_0}{{\omega_0}^2 - \omega^2} d\omega - \Re e[S(\omega_0)] \qquad (10)$$

where P stands for the Cauchy principal value. Actually, when an analytical formula for $\varepsilon(\omega)$ can be found for the IR part of the spectrum extending up to the large transparency window in the "visible" range, an approach combining the analytical integration for the IR range, and the one with discrete values for the UV range can be used. Indeed, the analytical modelling $\varepsilon_{IR}(\omega)$ valid in the IR range can nevertheless be defined on the whole spectrum (the quantities $\Im m[\varepsilon_{IR}(\omega)]$ and $\Im m[S_{IR}(\omega)]$, dropping down to zero for the visible-UV part of the spectrum), so that dividing the spectrum in two regions at an arbitrary cut point C located in the transparency region, one can rewrite (10) as :

$$r_a(\omega_0) = \frac{2}{\pi} P\int_0^C \Im m[S_{IR}(\omega)] \frac{\omega_0}{\omega_0^2 - \omega^2} d\omega - \Re e[S_{IR}(\omega_0)] + \frac{2}{\pi}\int_C^\infty \Im m[S(\omega)] \frac{\omega_0}{\omega_0^2 - \omega^2} d\omega \quad (11)$$

In (11), we have assumed $\omega_0$ to be in the IR range, so that $\Re e[S(\omega_0)] = \Re e[S_{IR}(\omega_0)]$, and in the last term, the condition $C>\omega_0$ enables one to remove the symbol for the "principal Cauchy value". Because we can extend $S_{IR}(\omega)$ to the visible-UV range (with $\Im m[S_{IR}(\omega)]$ taking a zero value), and using the equivalence between eqs. (10) and (3), one can introduce two separate contributions for $r_a(\omega_0)$ ,.

$$r_a(\omega_0) = \frac{2}{\pi}\int_0^\infty \frac{\omega_0}{\omega_0^2 + u^2} \frac{\varepsilon_{IR}(iu) - 1}{\varepsilon_{IR}(iu) + 1} du \quad + \quad \frac{2}{\pi}\int_C^\infty \Im m[S(\omega)] \frac{\omega_0}{\omega_0^2 - \omega^2} d\omega \qquad (12)$$

In eq. (12), the first term is hence easily evaluated through the analytical IR description of the permittivity, while the integration in the UV region can be performed numerically on the basis of discrete tabulated values. Moreover, the smoothing effect of the $\omega_0/(\omega_0^2-\omega^2)$ factor makes this second term in eq. (11) rather insensitive to an accurate knowledge of $\Im m[S(\omega)]$ in the UV region.

The overall smooth nature of $r_a(\omega)$ is illustrated in figure 5, where the non resonant dielectric coefficient is plotted for 5 materials of a large interest for our current experiments, and for which analytical formula in the IR range are easily found in the literature. This smooth behaviour justifies that we provide in Table 1 the values of $r_a(\omega)$ for only a selected number of frequencies (namely, 500, 1000, 25000, 10000cm$^{-1}$) . It is clear from such a smooth response that the spectroscopic accuracy of the $(n,\kappa)$ values is by far less critical than for the resonant contribution. Rather, only a very systematic error in the bulk material measurement could induce serious flaws on the $r_a(\omega)$ value. Inside this monotone decrease of $r_a(\omega)$, the UV contribution can change the dielectric image by a few %, especially in the vicinity of the transparency window (usually around the visible range). Figure 6 provides an illustration of the specific UV contribution for the case of $BaF_2$. In spite of its extremely broad transparency window, up to relatively deep UV, the specific UV contribution of $BaF_2$ reaches sizeable effects for wavelengths shorter than 1 µm. Note that the UV resonances makes the dielectric image coefficient finally decaying to zero, instead of reaching an asymptotic value resulting from $\varepsilon_{IR}(\infty)\neq 1$, and so *decreases* the value of $r_a(\omega)$ relatively to what would be calculated from the sole IR contributions ; moreover, far from the UV resonance, this decay is simply governed by the wing of the $\omega_0/(\omega-\omega_0)^2$ factor.

**6.Conclusion**

This work has been triggered by various uncertainties affecting theoretical predictions regarding our own experimental projects [3,5]. The critical analysis about the various data for YAG, and a specific reflectivity measurement, shows that our theoretical evaluation for $Cs(6D_{3/2})$ in front of a YAG window [3] is most probably to be revised. However, when the predictions for a given atom-surface system are sharply dependent upon the details of the surface resonance, the measurement of the bulk properties, and notably of reflectivity, should be operated in the operating temperature conditions. Even if index and absorption coefficients are usually not too dependent on the temperature, a tiny temperature change in the slope of the reflectivity response may indeed have an important consequence for the surface response. On more general grounds, the present results should be helpful if one needs to tailor for a given excited state, the atom-surface interaction. If our work has been here limited to an interaction with a planar surface, the extension to other shapes, including those of interest for nanotechnologies, should be straightforward if the surface response $S\equiv(\varepsilon-1)/(\varepsilon+1)$ is replaced by the adequate one. In particular, equivalent discussions on the influence of the uncertainties regarding the bulk measurements should still stand, as well as the influence of the UV transitions.

This work has been partially supported in the frame of the FASTNet project (European Union supported network HPRN-CT-2002-00304) and through the French-Bulgarian RILA cooperation. We acknowledge the contribution of F. Gervais (Tours University), and of I. Maurin (Paris13), for the YAG windows reflectivity measurements, of P. Todorov (Sofia) for checking files in table 1, and we thank Michèle Fichet (Paris13) for numerous discussions and her calculation for the fused silica.

**Table caption**

Table1: Characteristic amplitudes, and positions of the extreme amplitudes for the complex value surface response S(ω). The tabulated materials are alphabetically ordered materials. Only the main resonances are indicated, but some materials exhibit multiple resonances of a comparable size. To allow an approximate determination of the image coefficient, the value of $r_a(\omega_0)$ at fixed IR frequencies is also provided when IR analytical data is available. The values are italicized when the UV corrections are not taken into account, and appear in normal typing when data is available for the UV correction. The suffix"-bi" follows the name for a birefringent material : in such cases, one has taken $\varepsilon(\omega)=\sqrt{\varepsilon_o(\omega)\varepsilon_e(\omega)}$, as justified for a principal axis perpendicular to the window surface (see [10]). In the reference column, the reference to the Palik *ed.* Handbook [6] is simply indicated by the volume number, and first page of the chapter.

## Figure captions

Figure 1: Amplitude of the surface resonances for various materials, arranged as a function of frequency. Note the change of the vertical scale for the different sub-ranges. Each resonance is located on the peak value of $\Im m[S(\omega)]$. The dot indicates this peak value of $\Im m[S(\omega)]$, the vertical bar describes the peak-to-peak value of $2\Re e[S(\omega)]$ [the factor of 2 comes from eq.(5)]. For birefringent windows, one has taken: $\varepsilon(\omega)=\sqrt{\varepsilon_o(\omega)\varepsilon_e(\omega)}$.

Figure 2: Comparison of the optical constants and resonances for AlSb, according to two different bulk analytical determinations (dashed line[14], full line [15], the intermediate points -or squares for κ- correspond to the tabulated values given in [6]): (a) plot of the n and κ values ; (b) extrapolated (single) surface resonance for $\Im m[S(\omega)]$; (c) extrapolated (single) surface resonance for $\Re eS(\omega)$; (d) predicted reflectance according to the (n, κ) values.

Figure 3: Comparison of the surface resonances, and reflectance, for InSb, according to differing data (full line : [40], dashed line: [41], the intermediate points -or squares for κ- correspond to the tabulated values given in [6]) : (a) plot of the n and κ values ; (b) far IR plot of the n and κ values : (c) $\Re eS(\omega)$; (d) reflectance.

Figure 4: Comparison of the surface resonances, and reflectance, for YAG, according to the differing data for bulk optical constants of [36-38] ( as indicated) : (a) $\Re eS(\omega)$; (b) $\Im mS(\omega)$ ; (c) reflectance. In addition, (c) includes the reflectance measured - at room temperature- for the two windows of a vapour cell currently used in one of our experimental programme. The (not shown) reflectivity for a YAG powder [39] is in sensitive agreement with the calculated reflectivity derived from [36] or [38].

Figure 5 : $r_a(\omega_0)$ as a function of the frequency $\omega_0$ for some current types of windows (as indicated). The plot spans over two adjacent ranges of frequency. The calculation (based on bulk values given in [6]) takes into account the influence of the UV absorption band.

Figure 6 : The influence of the UV resonance on the $r_a(\omega_0)$ value for $BaF_2$. (a): plot of $\Im mS(\omega)$ spanning from the IR to the UV regions - note the change of frequency scale around

the transparency window- ; (b) : plot of $r_a(\omega_0)$ as obtained with and without taking into consideration of the remote UV resonances.

Saltiel et al, Table 1

| Material | frequency ($cm^{-1}$) for max. value of $\Im m(S)$ | frequency ($cm^{-1}$) for min. value of $\Re e(S)$ | frequency ($cm^{-1}$) for max value of $\Re e(S)$ | max. value of $\Im m(S)$ | max value of $\Re e(S)$ | min. value of $\Re e(S)$ | $r_a(\omega_0)$ for | | | | References |
|---|---|---|---|---|---|---|---|---|---|---|---|
| | | | | | | | 10000 $cm^{-1}$ | 2500 $cm^{-1}$ | 1000 $cm^{-1}$ | 500 $cm^{-1}$ | |
| **AgBr** | 125.5 | 113 | 138 | 1 | 1.17 | 0.167 | *0.64*<br>0.58 | *0.65*<br>0.63 | 0.66<br>0.65 | 0.67<br>0.67 | III-553 |
| **AgCl** | 184 | 170.6 | 197 | 1.12 | 1.22 | 0.001 | *0.60*<br>0.55 | *0.61*<br>0.60 | *0.63*<br>0.62 | *0.65*<br>0.65 | III-553 |
| **$AgGaS_2$** | 237 | 225 | 249 | 0.46 | 1.03 | 0.58 | | | | | III-573 |
| **$AgGaSe_2$** | 270 | 261 | 270 | 1.48 | 1.4 | -0.22 | | | | | III-573 |
| **AgI** | 116 | 97 | 135 | 0.256 | 0.81 | 0.55 | *0.66* | *0.66* | *0.67* | *0.67* | III-553[(a)] |
| **AgI** (data) | 113 | 104 | 124 | 0.49 | 0.83 | 0.31 | | | | | III-553[(b)] |
| **$Al_2O_3$** | 820 | 810 | 830 | 11.48 | 6.35 | -5.17 | *0.53*<br>0.50 | *0.58*<br>0.58 | *0.64*<br>0.64 | *0.70*<br>0.70 | II-761, III-653, [11-13] |
| **AlAs** | 397 | 393 | 401 | 1.84 | 1.71 | -0.13 | *0.78* | *0.79* | *0.79* | *0.80* | II-489 |
| **$Al_xGa_{1-x}As$** x=0.3 | 277 | 274 | 280 | 0.56 | 1.11 | 0.57 | *0.82* | *0.82* | *0.83* | *0.83* | II-513 |
| **$Al_xGa_{1-x}As$** x=0.3 | 377 | 372 | 382 | 0.68 | 1.12 | 0.44 | *0.82* | *0.82* | *0.83* | *0.83* | II-513 |
| **ALON** | 850 | 814 | 956 | 1.89 | 1.45 | -0.37 | *0.54* | *0.59* | *0.65* | *0.70* | II-777 |
| **AlSb** | 337.7 | 336 | 338.7 | 3.57 | 2.07 | -0.52 | *0.82* | *0.82* | *0.82* | *0.82* | II-501, [14,15] |
| **$As_2S_3$**-fused | 354 | 317.5 | 377.4 | 0.37 | 0.94 | 0.49 | | | | | I-641 |
| **$As_2Se_3$**-bi | 234.9 | 229.8 | 240 | 0.5 | 1.07 | 0.56 | | | | | I-623 |
| **$As_2Se_3$**-fused | 232 | 220.1 | 268.7 | 0.23 | 0.92 | 0.67 | | | | | I-623 |
| **$BaF_2$** | 333 | 312 | 350 | 3.83 | 2.41 | -1.71 | *0.38*<br>0.36 | *0.41*<br>0.41 | *0.46*<br>0.46 | *0.52*<br>0.52 | III-683(c), [16,17] |
| **$BaMg_{1/3}Nb_{2/3}O_3$** | 441 | 418 | 450 | 1.84 | 1.93 | -0.23 | *0.63* | *0.65* | *0.70* | *0.74* | [18] |
| **$BaMg_{1/3}Nb_{2/3}O_3$** | 708 | 683 | 721 | 1.71 | 1.26 | -0.33 | *0.63* | *0.65* | *0.70* | *0.74* | [18] |
| **BN**-cubic | 1292 | 1272 | 1312 | 3.72 | 2.53 | -1.2 | *0.65* | *0.67* | *0.70* | *0.72* | III-425 |
| **$BaTiO_3$** -bi | 463 | 459.5 | 466 | 2.32 | 2.29 | 0.21 | *0.70*<br>0.63 | *0.75*<br>0.73 | *0.81*<br>0.80 | *0.86*<br>0.86 | II-789 ,[19] |
| **$BaTiO_3$** -bi | 665 | 641 | 685 | 3.78 | 2.28 | -1.51 | *0.70*<br>0.63 | *0.75*<br>0.73 | *0.81*<br>0.80 | *0.86*<br>0.86 | II-789, [19] |
| **BeO-bi** | 1013 | 1007 | 1020 | 21.26 | 11.28 | -10.11 | *0.52* | *0.57* | *0.63* | *0.67* | II-805 |
| **BeO**-ceramics. | 1007 | 985 | 1022 | 6.79 | 3.56 | -2.92 | | | | | II-805 |
| **$Bi_{12}GeO_{20}$** | 365 | 345 | 375 | 1.19 | 1.92 | 0.01 | | | | | III-403 |
| **$Bi_{12}GeO_{20}$** | 490 | 480 | 495 | 1.11 | 1.33 | 0.05 | | | | | III-403 |
| **$Bi_{12}GeO_{20}$** | 545 | 538 | 550 | 1.59 | 1.27 | -0.92 | | | | | III-403 |
| **$Bi_{12}GeO_{20}$** | 695 | 690 | 700 | 1 | 1.35 | -0.31 | | | | | III-403 |
| **$Bi_{12}SiO_{20}$** | 357 | 353 | 366 | 1.74 | 1.49 | -0.89 | | | | | III-403 |
| **$Bi_{12}SiO_{20}$** | 495 | 491 | 502 | 1.93 | 1.62 | -0.46 | | | | | III-403 |
| **$CaCO_3$** | 362 | 357 | 373 | 4.55 | 3.26 | -2.69 | | | | | III-701 |
| **$CaF_2$** | 417 | 401 | 432 | 5.55 | 3.1 | 2.44 | *0.36*<br>0.34 | *0.39*<br>0.39 | *0.46*<br>0.46 | *0.52*<br>0.52 | II-815, [16] |
| **$CaMoO_4$** -bi | 876 | 872 | 880 | 7.93 | 4.57 | -3.34 | *0.59* | *0.62* | *0.65* | *0.69* | [20] |
| **$CaWO_4$** -bi | 873 | 869 | 876 | 10.7 | 5.91 | -4.83 | *0.56* | *0.59* | *0.64* | *0.68* | [20] |
| **$CdGeAs_2$** -bi | 209 | 204 | 213.5 | 0.14 | 0.96 | 0.82 | *0.88*<br>0.80 | *0.88*<br>0.86 | *0.88*<br>0.87 | *0.88*<br>0.88 | III, p.445 |
| **$CdGeAs_2$** -bi | 281 | 278 | 283.5 | 0.48 | 1.12 | 0.614 | *0.88*<br>0.80 | *0.88*<br>0.86 | *0.88*<br>0.87 | *0.88*<br>0.88 | III, p.445 |
| **CdS -**bi | 293 | 291 | 295.2 | 6.69 | 4.1 | -2.62 | *0.69*<br>0.58 | *0.70*<br>0.67 | *0.71*<br>0.70 | *0.72*<br>0.72 | II, p.579 |
| **CdSe** | 204 | 202 | 205 | 4.13 | 3.6 | -2.1 | | | | | II, p.559 |
| **CdTe** | 166 | 162.5 | 169.3 | 1.73 | 1.63 | -0.1 | *0.75* | *0.76* | *0.76* | *0.77* | I, p.409 |
| **CsBr** | 106 | 94 | 111 | 2.11 | 1.66 | -0.77 | | | | | III, p.717 |
| **CsCl** | 150 | 145.1 | 157 | 3.12 | 2.46 | -0.8 | | | | | III, p.731 |
| **CsI** | 85 | 78.2 | 95 | 1.26 | 1.52 | -0.15 | | | | | II, p.853 |
| **$Cu_2O$** | 149 | 148.8 | 150 | 0.53 | 1.24 | 0.33 | | | | | II, p.875 |
| **$CuGaS_2$** -bi | 381 | 379 | 382 | 2.99 | 2.38 | -0.46 | *0.72*<br>0.63 | *0.73*<br>0.71 | *0.74*<br>0.73 | *0.74*<br>0.74 | III, p.459 |
| **$FeS_2$** | 437.8 | 435 | 441 | 1.06 | 1.43 | 0.38 | *0.91*<br>0.74 | *0.91*<br>0.88 | *0.91*<br>0.90 | *0.92*<br>0.91 | III, p.507 |
| **GaAs** | 290 | 289 | 291.5 | 2.95 | 2.28 | -0.6 | *0.83*<br>0.75 | *0.84*<br>0.82 | *0.84*<br>0.83 | *0.84*<br>0.84 | I-409, [21] (d) |
| **GaP** | 399.5 | 398 | 402.5 | 5.37 | 2.86 | -1.3 | *0.74*<br>0.67 | *0.75*<br>0.73 | *0.75*<br>0.75 | *0.76*<br>0.76 | I-445, III-38, [22] |
| **GaSb** | 240.4 | 239 | 248 | 1.41 | 1.52 | 0.11 | *0.83* | *0.83* | *0.84* | *0.84* | II- 597 |
| **GaSe** -bi | 245 | 244 | 255 | 1.67 | 1.82 | -0.03 | *0.74* | *0.76* | *0.77* | *0.78* | III-473 |
| **HAFNIA** | 642 | 603 | 684 | 1.72 | 1.45 | -0.21 | *0.63* | *0.67* | *0.72* | *0.76* | [23] |
| **HgCdTe** | 163 | 162 | 166 | 0.94 | 1.3 | 0.39 | *0.88* | *0.88* | *0.89* | *0.89* | II-665 |
| **InAs** | 238 | 236 | 240 | 1.53 | 1.61 | 0.083 | *0.84*<br>0.74 | *0.84*<br>0.82 | *0.85*<br>0.84 | *0.85*<br>0.85 | I-479 |
| **InP** | 341 | 339 | 343 | 3.85 | 2.73 | -1.1 | *0.81*<br>0.71 | *0.82*<br>0.79 | *0.82*<br>0.81 | *0.83*<br>0.82 | I-503 |

| | | | | | | | | | | |
|---|---|---|---|---|---|---|---|---|---|---|
| **InSb** | 73 | 68 | 78.5 | 0.71 | 1.28 | 0.57 | *0.88*<br>0.77 | *0.88*<br>0.86 | *0.89*<br>0.88 | *0.90*<br>0.89 | I-491[e] |
| **InSb** | 192 | 190 | 193.5 | 1.13 | 1.44 | 0.31 | *0.88*<br>0.77 | *0.88*<br>0.86 | *0.89*<br>0.88 | *0.90*<br>0.89 | I-491[e] |
| **KBr** | 142 | 138 | 150 | 2.51 | 2.2 | -0.45 | | | | | II-989 |
| **KCl** | 181 | 172 | 194 | 3.22 | 2.44 | -0.8 | | | | | I,-703 |
| **KI** | 122 | 119 | 128 | 2.26 | 2.02 | -0.13 | *0.46*<br>0.42 | *0.47*<br>0.46 | *0.49*<br>0.48 | *0.51*<br>0.50 | III-807[f], [24] |
| **$Li_2CaHfF_8$** -bi | 522 | 505 | 539 | 3.68 | 2.11 | -1.46 | *0.42* | *0.45* | *0.50* | *0.56* | [25] |
| **LiF** | 576 | 557 | 592 | 7.1 | 3.48 | -3.46 | *0.35*<br>0.33 | *0.41*<br>0.41 | *0.50*<br>0.49 | *0.57*<br>0.57 | I-675, [26] |
| **$LiIO_3$** -bi | 434 | 417 | 451 | 2.19 | 1.69 | -0.5 | *0.51* | *0.54* | *0.58* | *0.62* | [27] |
| **$LiIO_3$** -bi | 818 | 814 | 821 | 3.75 | 2.58 | -0.83 | *0.51* | *0.54* | *0.58* | *0.62* | [27] |
| **$LiNbO_3$** -bi | 828 | 812 | 844 | 4.71 | 3.01 | -1.67 | *0.67* | *0.71* | *0.76* | *0.81* | [28] |
| **$LiTaO_3$** -bi | 808 | 798 | 819 | 7.37 | 4.2 | -3.1 | *0.66*<br>0.61 | *0.70*<br>0.69 | *0.76*<br>0.75 | *0.81*<br>0.80 | [29] |
| **$MgAl_2O_4$-spinel** | 793 | 774 | 832 | 3.53 | 2.5 | -1.16 | *0.53* | *0.58* | *0.63* | *0.68* | II-883 |
| **$MgF_2$**-bi | 556 | 541.5 | 571 | 6.28 | 3.68 | -2.97 | *0.33*<br>0.31 | *0.37*<br>0.36 | *0.43*<br>0.43 | *0.49*<br>0.49 | II-899, [30] |
| **MgO** | 661 | 628 | 679 | 5.94 | 3.66 | -3 | *0.51*<br>0.48 | *0.56*<br>0.55 | *0.62*<br>0.62 | *0.68*<br>0.67 | II-919, [26] |
| **NaCl** | 222 | 217 | 228 | 3.46 | 2.7 | -0.51 | | | | | I-775 |
| **NaF** | 365 | 338 | 390 | 6 | 3.39 | -1.84 | | | | | II-1021 |
| **$NaNO_3$**-bi | 89 | 85.5 | 98 | 1 | 0.88 | 0.1 | *0.36* | *0.38* | *0.40* | *0.43* | III-871 |
| **$NaNO_3$**-bi | 245 | 234 | 258 | 1.85 | 1.48 | -0.353 | *0.36* | *0.38* | *0.40* | *0.43* | III,871 |
| **PbS** | 207 | 202 | 214 | 1.33 | 1.61 | 0.19 | | | | | I-525 |
| **PbSe** | 220 | 200 | 240 | 0.54 | 1.2 | 0.67 | | | | | I-517 |
| **$Pb_{1-x}Sn_xTe$**<br>x=0.21 | 120.5 | 114 | 126 | 0.46 | 1.19 | 0.73 | *0.95* | *0.96* | *0.96* | *0.96* | II-637 |
| **PbTe** | 252 | 240 | 263.5 | 0.67 | 1.29 | 0.621 | *0.94* | *0.95* | *0.95* | *0.96* | I-535 |
| **$PbWO_4$** -bi | 849 | 842 | 856 | 4.12 | 2.1 | -1.76 | *0.58* | *0.61* | *0.64* | *0.68* | [31] |
| **$PbZrO_3$** | 378 | 361 | 389 | 1.17 | 1.31 | 0.28 | *0.67* | *0.71* | *0.76* | *0.81* | [32] |
| **$PbZrO_3$** | 621.5 | 595.5 | 648 | 1.69 | 1.44 | -0.23 | *0.67* | *0.71* | *0.76* | *0.81* | [32] |
| **RbBr** | 112 | 106 | 117 | 1.88 | 1.79 | -0.36 | | | | | III-845 |
| **RbI** | 93 | 90 | 97 | 2.38 | 2.26 | -0.18 | | | | | III-857 |
| **Se** -bi | 104 | 100 | 107.5 | 0.06 | 0.84 | 0.78 | | | | | II-691 |
| **Se** -bi | 143 | 138.5 | 147 | 0.1 | 0.84 | 0.75 | | | | | II-691 |
| (N-doped) **Si** | 267 | 151.5 | 380 | 0.23 | 1.01 | 0.76 | | | | | I-547 |
| **SiC**-hexagonal | 948 | 945.5 | 950.5 | 15.53 | 8.52 | -7 | *0.75*<br>0.69 | *0.76*<br>0.75 | *0.78*<br>0.77 | *0.79*<br>0.79 | I-587 |
| **$SiO_2$**-bi | 1180 | 1177 | 1184 | 16.2 | 8.09 | -7.94 | *0.39*<br>0.37 | *0.45*<br>0.45 | *0.50*<br>0.50 | *0.54*<br>0.54 | I-719, [27] |
| **Silica** | 494 | 485 | 510 | 1.33 | 1.35 | 0.04 | *0.36*<br>0.34 | *0.42*<br>0.41 | *0.47*<br>0.46 | *0.51*<br>0.51 | I-749[(g)] |
| **Silica** | 1165 | 1133 | 1235 | 2.05 | 1.93 | -0.86 | *0.36*<br>0.34 | *0.42*<br>0.41 | *0.47*<br>0.46 | *0.51*<br>0.51 | I-749[(g)] |
| **$SrF_2$** | 335 | 325.5 | 343 | 7.03 | 4.6 | -3.28 | *0.36*<br>0.35 | *0.39*<br>0.39 | *0.44*<br>0.44 | *0.50*<br>0.50 | III-883, [16] |
| **$SrTiO_3$** | 462 | 460 | 470 | 3.84 | 5.62 | -1.13 | *0.70* | *0.75* | *0.81* | *0.86* | II-1035 , [33] |
| **$SrTiO_3$** | 736 | 720 | 748 | 6.76 | 3.96 | -3.16 | *0.70* | *0.75* | *0.81* | *0.86* | II-1035, [33] |
| **Te** -bi | 104 | 96 | 106 | 0.19 | 1.04 | 0.8 | *0.94*<br>0.82 | *0.94*<br>0.92 | *0.94*<br>0.93 | *0.95*<br>0.94 | II-709 |
| **$TeO_2$** -bi | 374 | 371 | 377.5 | 1.69 | 2.01 | 0.64 | *0.70* | *0.73* | *0.77* | *0.80* | [34] |
| **$TeO_2$**-bi | 716 | 709.5 | 725 | 2.14 | 2.2 | 0.46 | *0.70* | *0.73* | *0.77* | *0.80* | [34] |
| **$TeO_2$** -bi | 775 | 767 | 811 | 1.17 | 0.68 | -0.25 | *0.70* | *0.73* | *0.77* | *0.80* | [34] |
| **$TiO_2$** -bi | 764 | 743 | 785 | 3.54 | 2.55 | -0.9 | *0.76* | *0.80* | *0.85* | *0.89* | [35] |
| **TlBr** -bi | 108 | 101 | 116 | 1.86 | 1.79 | -0.07 | | | | | III-923 |
| **TlBr** -bi | 107.5 | 96 | 114 | 1.6 | 1.47 | -0.2 | | | | | III-923 |
| **TlCl** | 159 | 153 | 166 | 2.8 | 2.31 | -0.66 | | | | | III-923 |
| **TlClBr (KRS6)** | 139 | 123 | 159 | 1.13 | 1.43 | 0.29 | | | | | III-923 |
| **TlI** | 87.5 | 82 | 97.5 | 1.42 | 1.75 | 0.13 | | | | | III-923 |
| **$Y_2O_3$** (yttria) | 526 | 521 | 530 | 6.11 | 4.48 | -1.43 | *0.58*<br>0.53 | *0.61*<br>0.60 | *0.66*<br>0.65 | *0.70*<br>0.70 | II-1079 |
| **$Y_2O_3$** (yttria) | 586 | 581 | 591 | 8.27 | 3.99 | -4.14 | *0.58*<br>0.53 | *0.61*<br>0.60 | *0.66*<br>0.65 | *0.70*<br>0.70 | II-1079 |
| **YAG** | 540 | 535 | 545 | 3.1 | 2.4 | -0.22 | *0.56*<br>0.52 | *0.60*<br>0.59 | *0.65*<br>0.65 | 0.69<br>0.69 | III-963(h) [36-39] |
| **YAG** | 576 | 572 | 581 | 2.03 | 1.3 | -0.52 | *0.56*<br>0.52 | *0.60*<br>0.59 | *0.65*<br>0.65 | 0.69<br>0.69 | III-963(h) [36-39] |
| **YAG** | 823.5 | 815 | 832 | 5.6 | 2.62 | -2.3 | *0.56*<br>0.52 | *0.60*<br>0.59 | *0.65*<br>0.65 | 0.69<br>0.69 | III-963(h) [36-39] |
| **YGG** | 470 | 466 | 473 | 3.54 | 2.93 | -0.34 | *0.57* | *0.60* | *0.65* | *0.69* | [38] |
| **YGG** | 491 | 488 | 495 | 3.95 | 2.29 | -1.44 | *0.57* | *0.60* | *0.65* | *0.69* | [38] |
| **YGG** | 673.5 | 668 | 678.5 | 3.87 | 3 | -0.33 | *0.57* | *0.60* | *0.65* | *0.69* | [38] |
| **YGG** | 700.5 | 695.5 | 706 | 4.36 | 1.6 | -2.3 | *0.57* | *0.60* | *0.65* | *0.69* | [38] |
| **YIG** | 428 | 422 | 435 | 2.98 | 2.4 | -0.56 | *0.67* | *0.70* | *0.74* | *0.78* | [38] |
| **YIG** | 690 | 681 | 699.5 | 3.66 | 2.31 | -1.28 | *0.67* | *0.70* | *0.74* | *0.78* | [38] |

Saltiel et al, Table 1

| | | | | | | | | | | | |
|---|---|---|---|---|---|---|---|---|---|---|---|
| **$ZnGeP_2$** -bi | 402 | 401 | 411 | 0.78 | 1.2 | 0.44 | *0.82*<br>0.70 | *0.82*<br>0.79 | *0.82*<br>0.81 | *0.83*<br>0.82 | III-637 |
| **ZnN** | 24.5 | 22 | 27 | 1.39 | 1.62 | 0.28 | | | | | III-351 |
| **ZnS** | 342.5 | 339 | 346 | 4.86 | 3.16 | -1.7 | *0.70* | *0.71* | *0.73* | *0.74* | I-597 |
| **ZnSe** | 247 | 243 | 250 | 3.26 | 2.36 | -0.9 | *0.69*<br>0.59 | *0.69*<br>0.67 | *0.70*<br>0.69 | *0.71*<br>0.71 | II-737 |
| **ZnTe** | 202 | 200 | 203.5 | 4.03 | 2.78 | -1.25 | *0.74*<br>0.65 | *0.74*<br>0.72 | *0.75*<br>0.74 | *0.76*<br>0.75 | II-737 |

(a) the table values are obtained on the base of analytical formula for $\varepsilon(\omega)$

(b) the table values are obtained on the base of numerical table for $n(\omega)$ and $\kappa(\omega)$ given in [6].

(c) The (n,κ) data of [6] is derived from the Hoffman fitting curve [17], rather than from the Kaiser curve [16] , because Hoffmann introduced additional terms in order to take into account the impurities and defects of the real material, one of the modes taking into account the two-photon absorption.

(d) Note that in [6], there is a mistakes in numerical table for $n(\omega)$ and $\kappa(\omega)$ in the decimal point position - clearly visible through a plotting.

(e) Note that in [6], there is error in the sign in the formula for ε in the denominator of second term that accounts for the free electrons.

(f) Note that in [6], there is misprint for the value of one of the second resonance – should be $\omega_2 = 144$ cm$^{-1}$ [24]

(g) For this material all table values including $r_a(\omega_o)$ are found on the base of numerical table for $n(\omega)$ and $\kappa(\omega)$ given in [6]. For $r_a(\omega_o)$, eq. (10) has been used.

(h) Note that in [6], there is a mixing between the references of the columns so that it does not allow a correct plot of $\Re$e(S) and $\Im$m(S) starting from the $n(\omega)$ and $\kappa(\omega)$ values.

Saltiel et al, figure 1

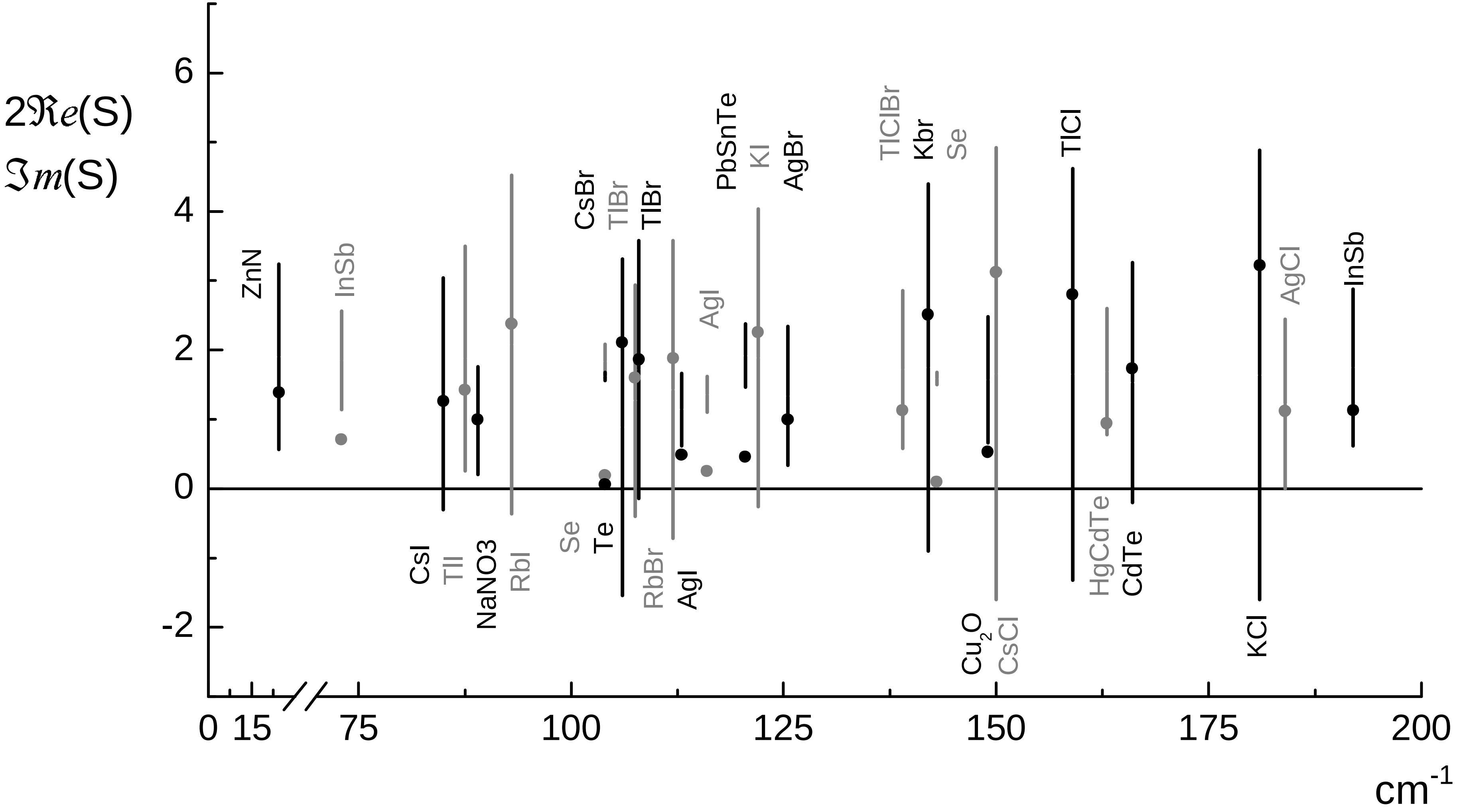

Saltiel et al, figure 1

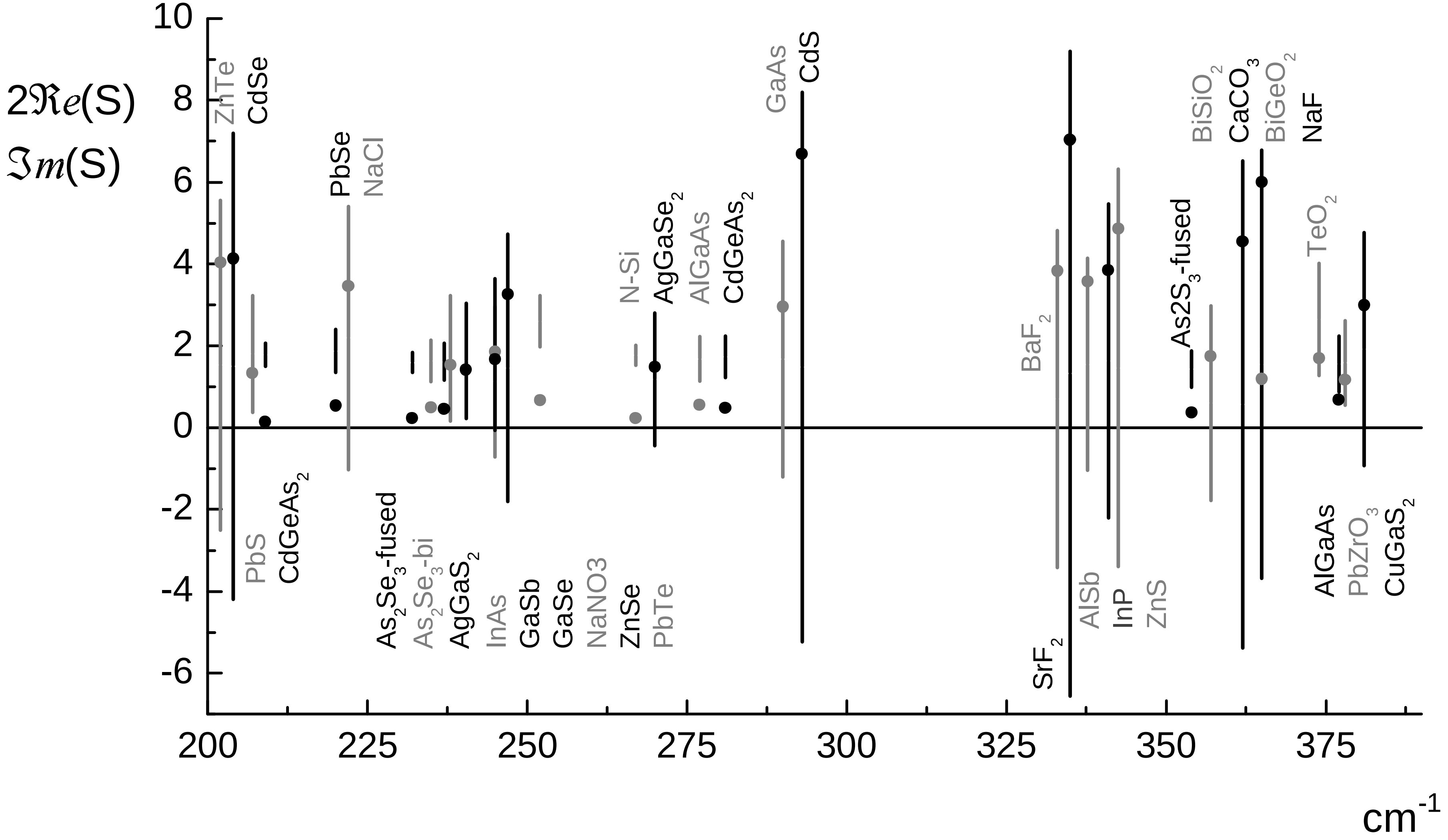

Saltiel et al, figure 1

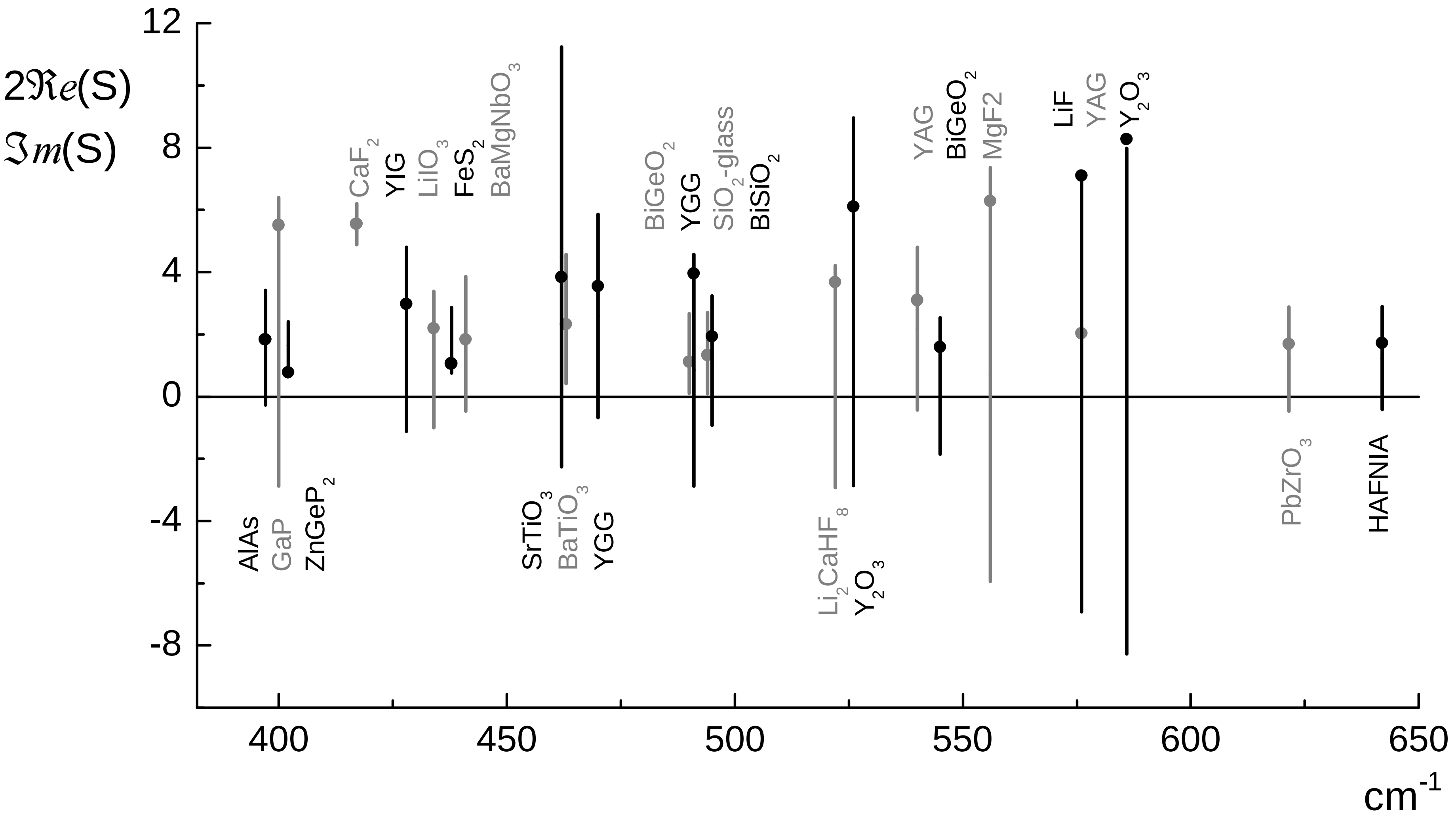

Saltiel et al, figure 1

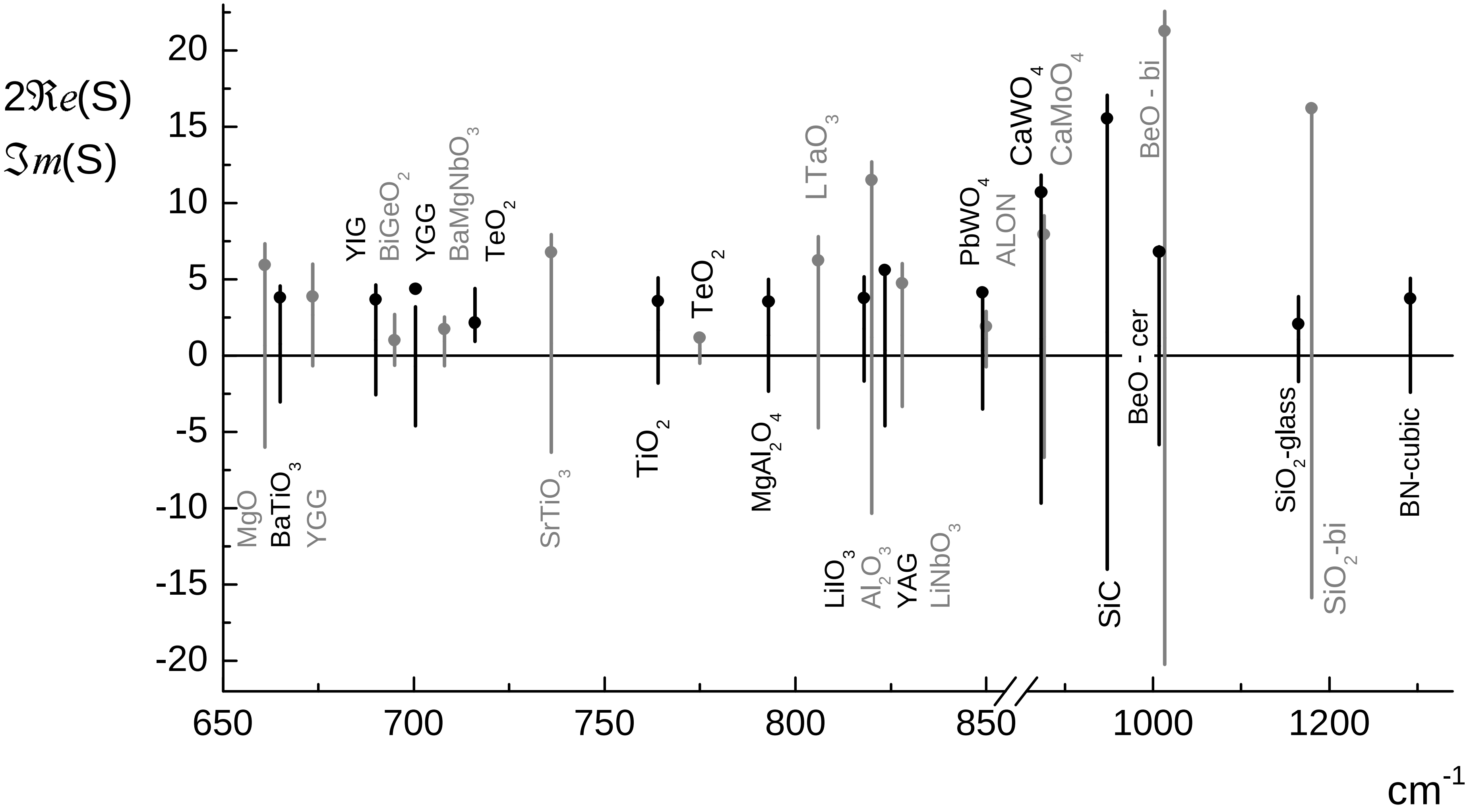

Saltiel et al, figure 2

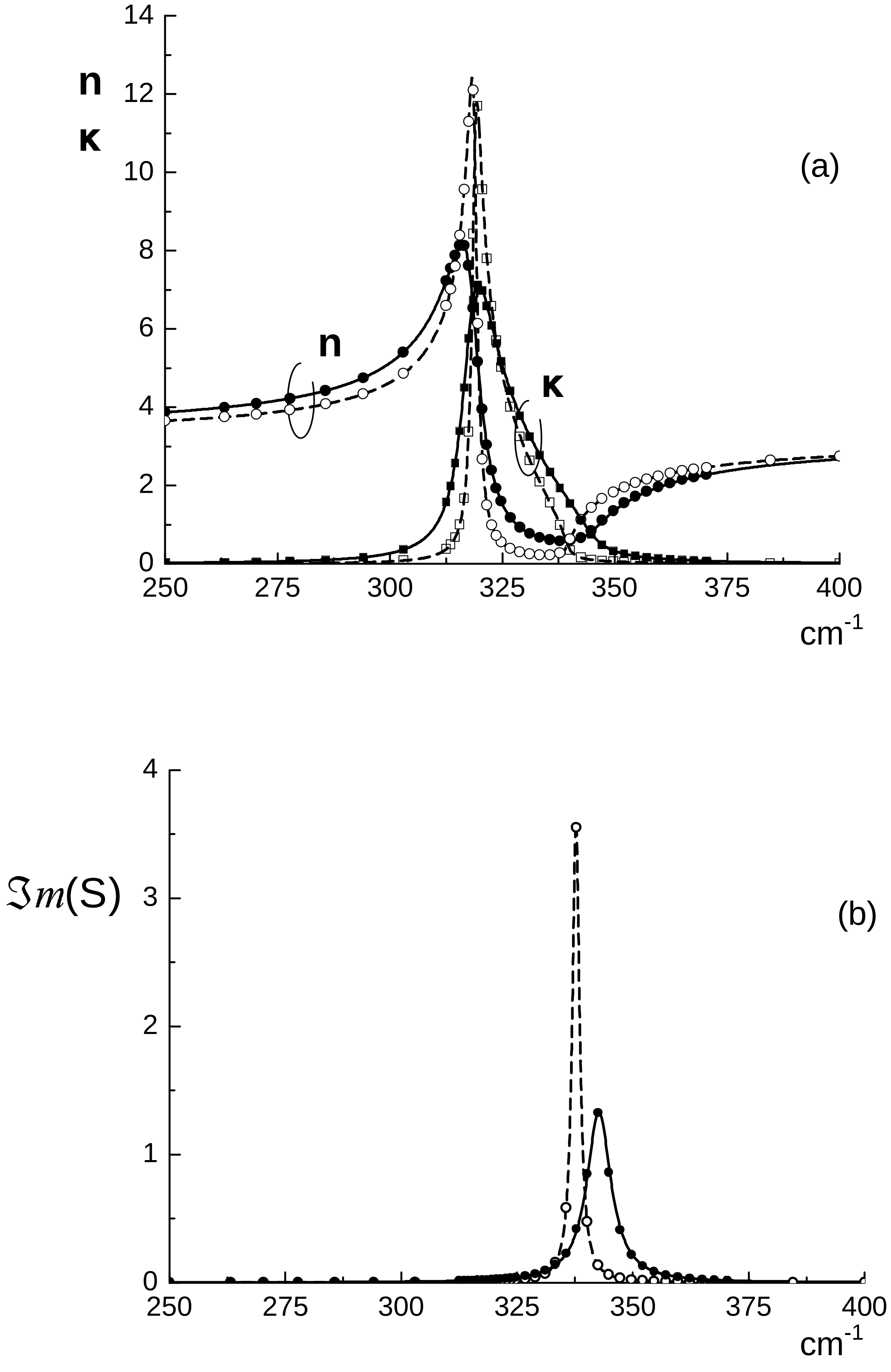

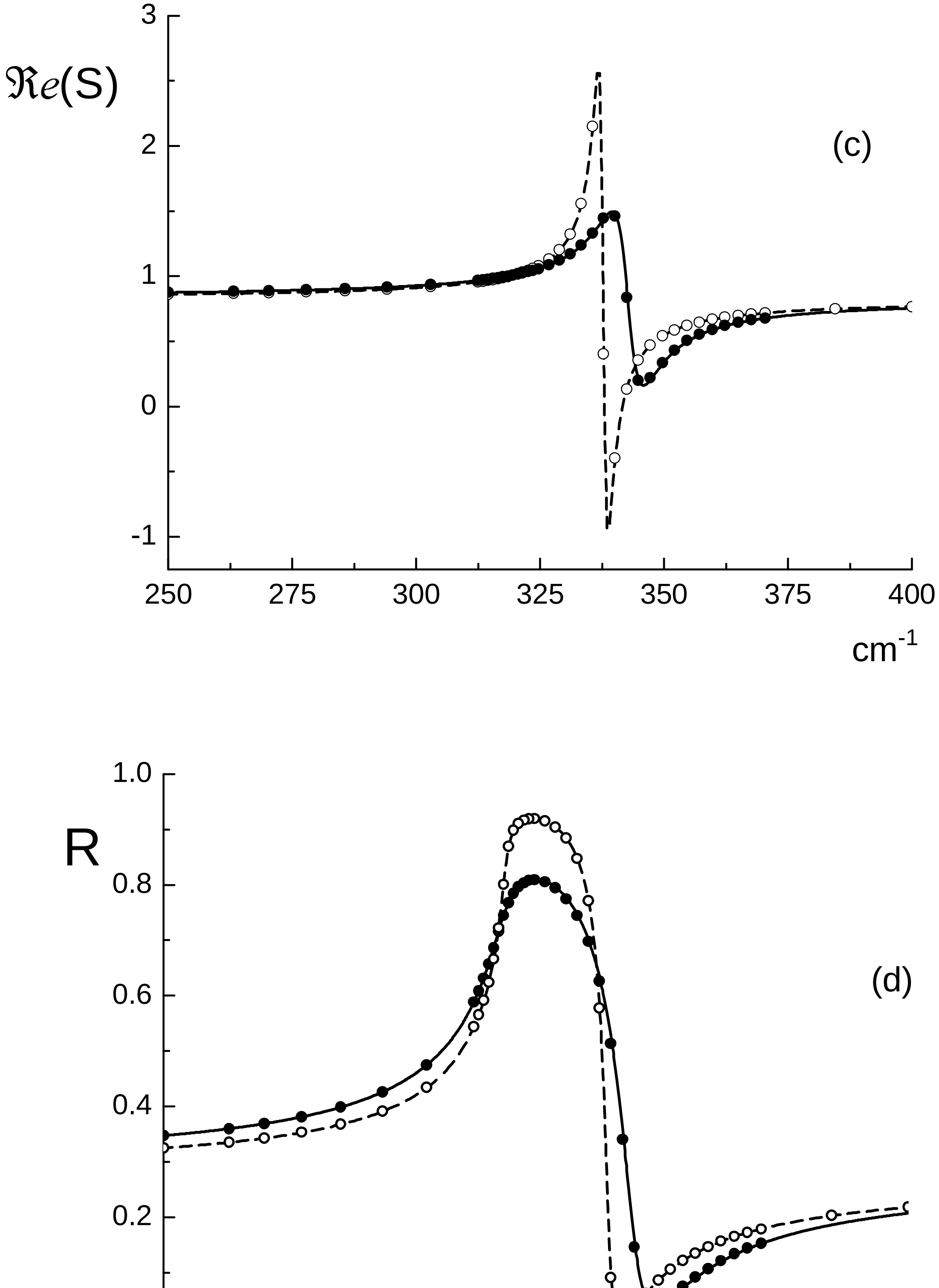

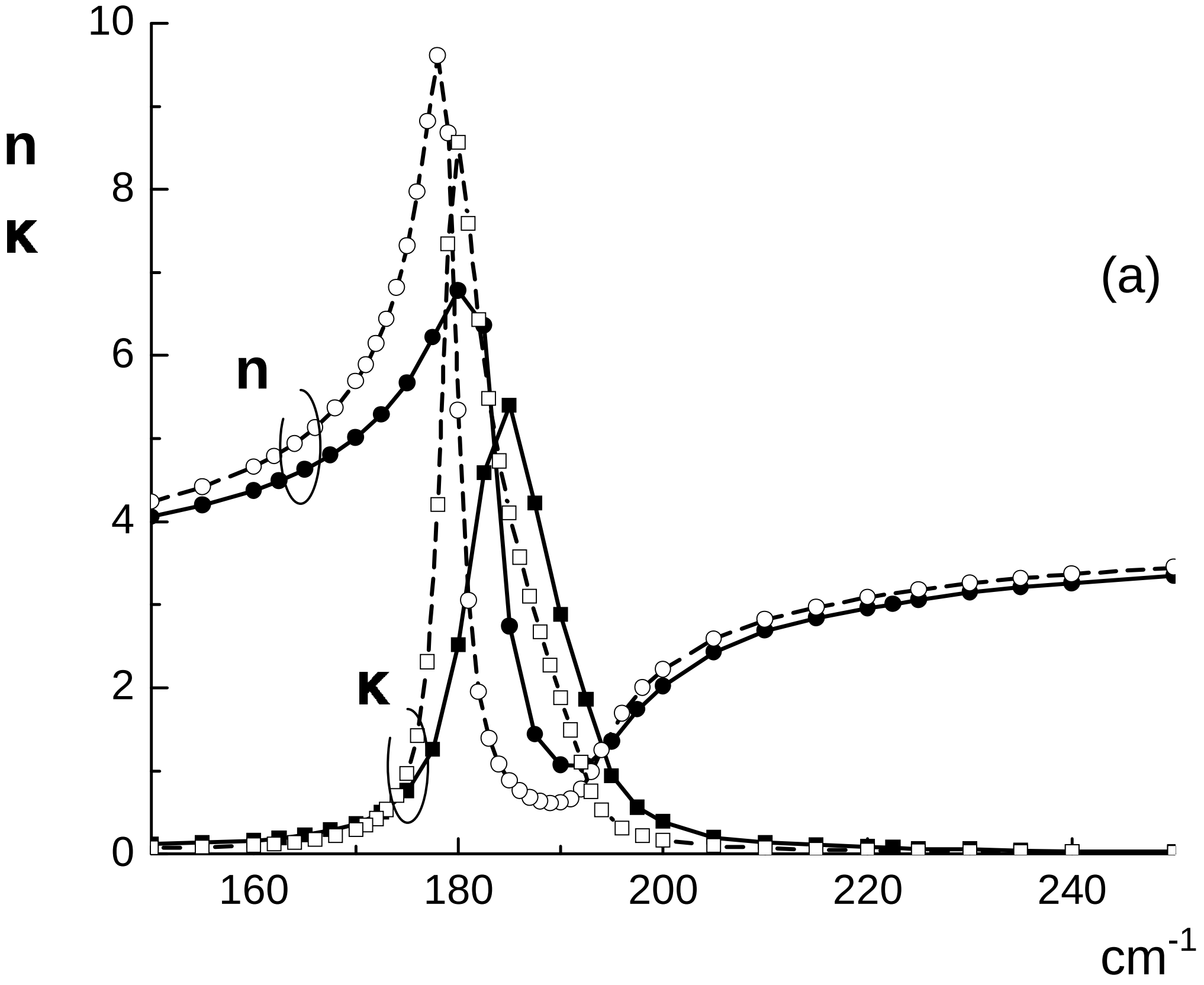

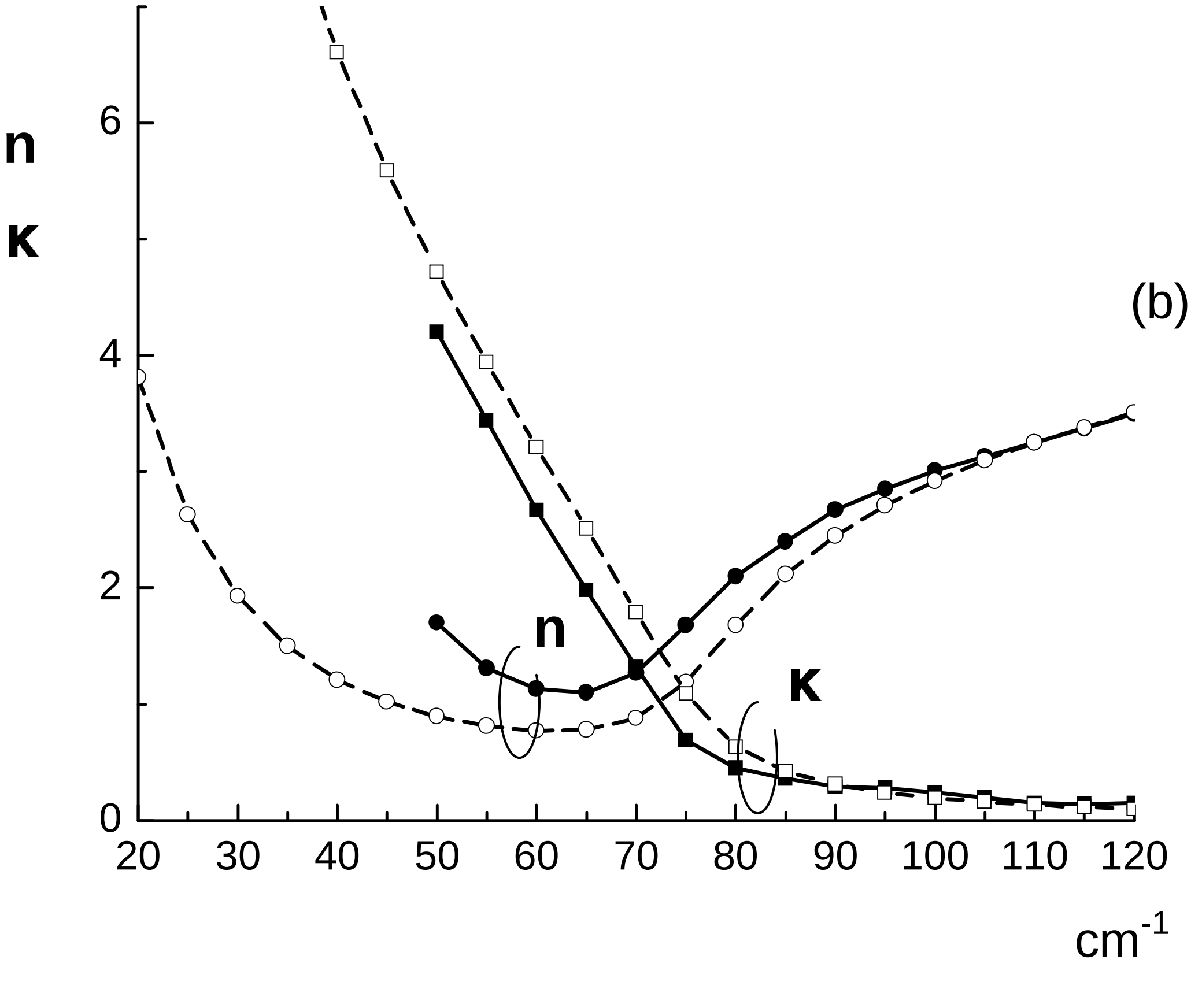

Saltiel et al, figure 3

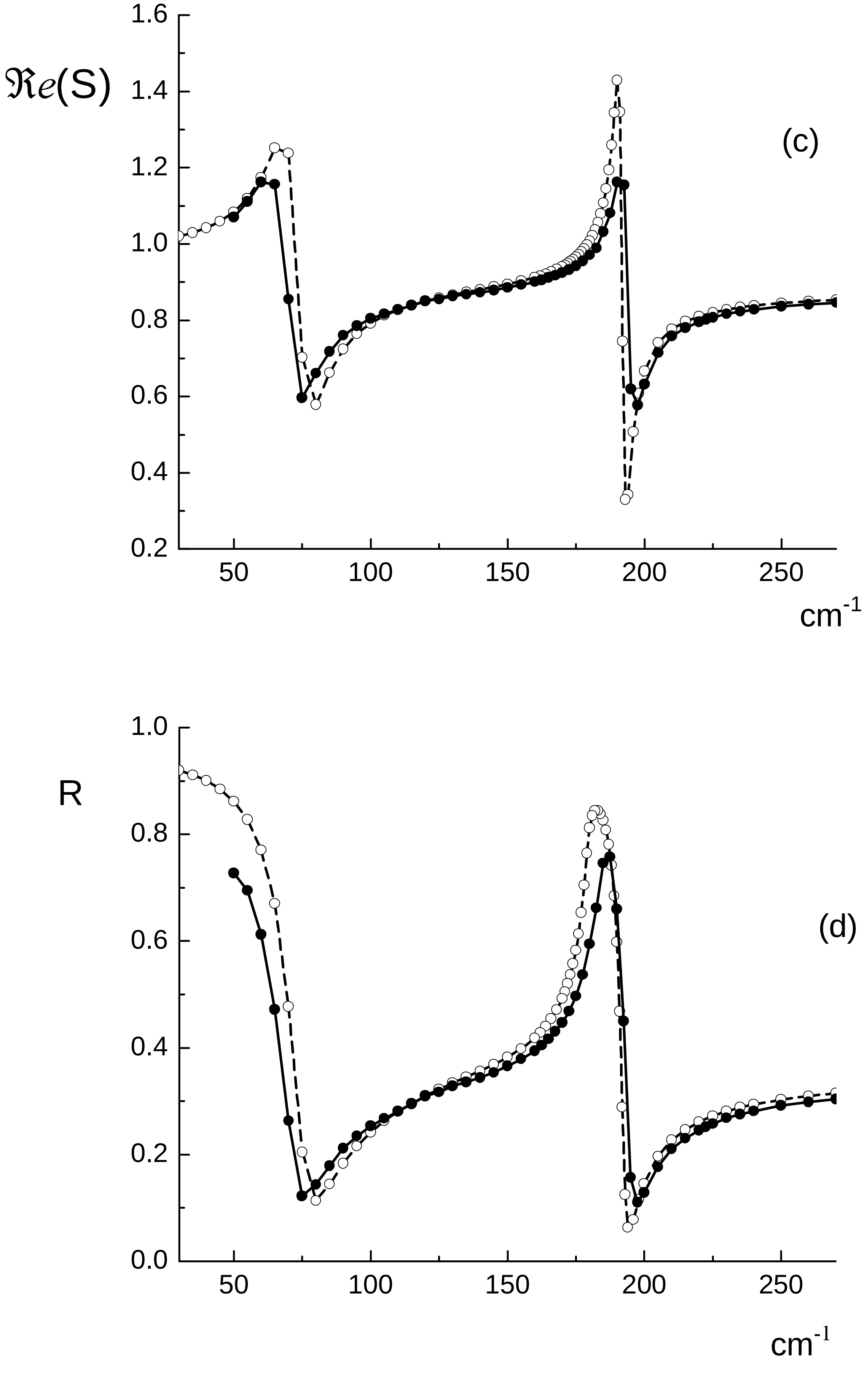

Saltiel et al, figure 3

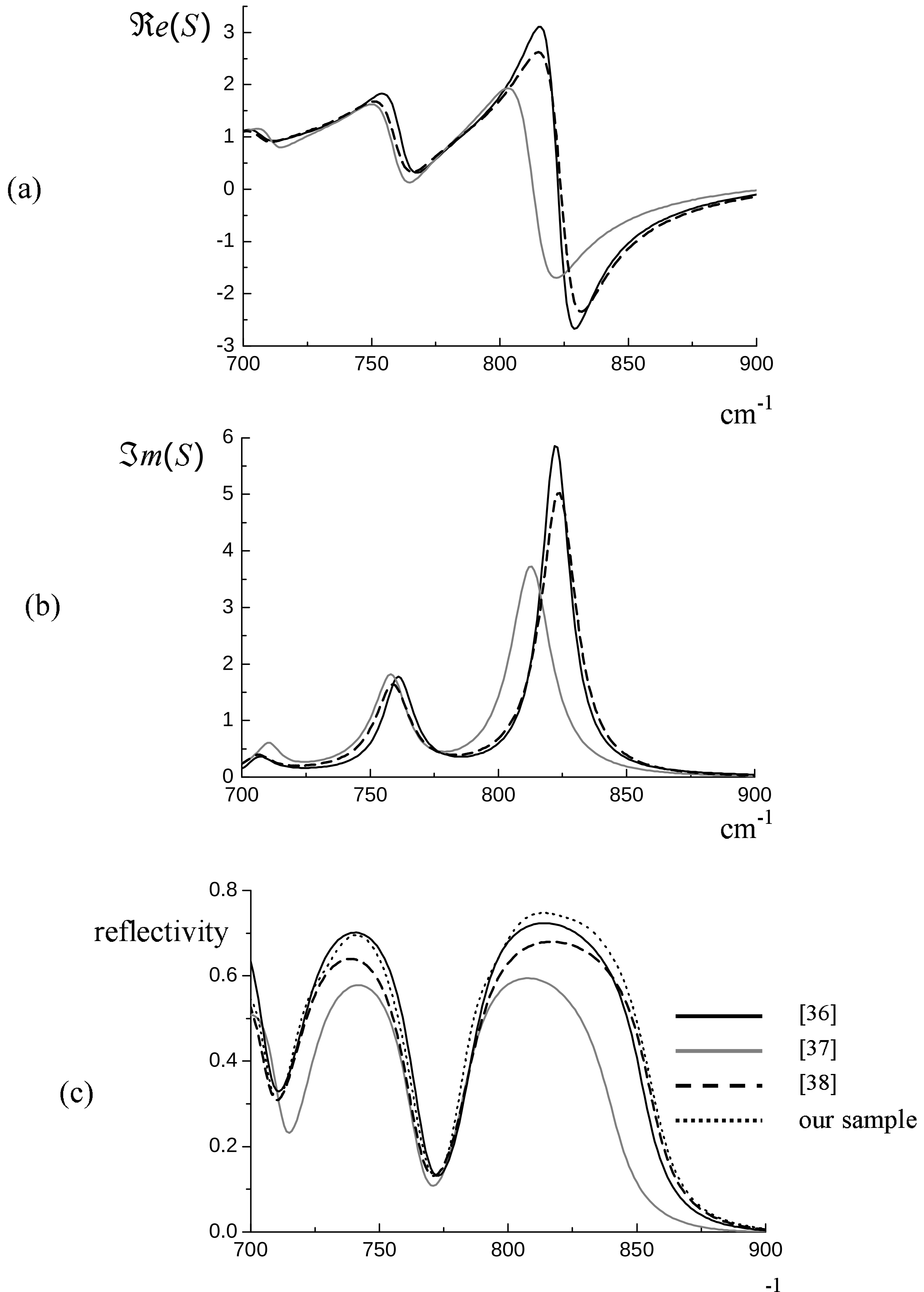

Saltiel et al, figure 4

Fig 4, in the text is Figure 5

$\Re e(S)$ $\Im e(S)$

$\Re e(S)$ *and* $\Im m(S)$

[20]

[17]

[21]

our sample

[20]

[17]

[21]

our sample

frequency, cm-1

$cm^{-1}$

$\Im m(S)$

Saltiel et al, figure 4

Saltiel et al, figure 5

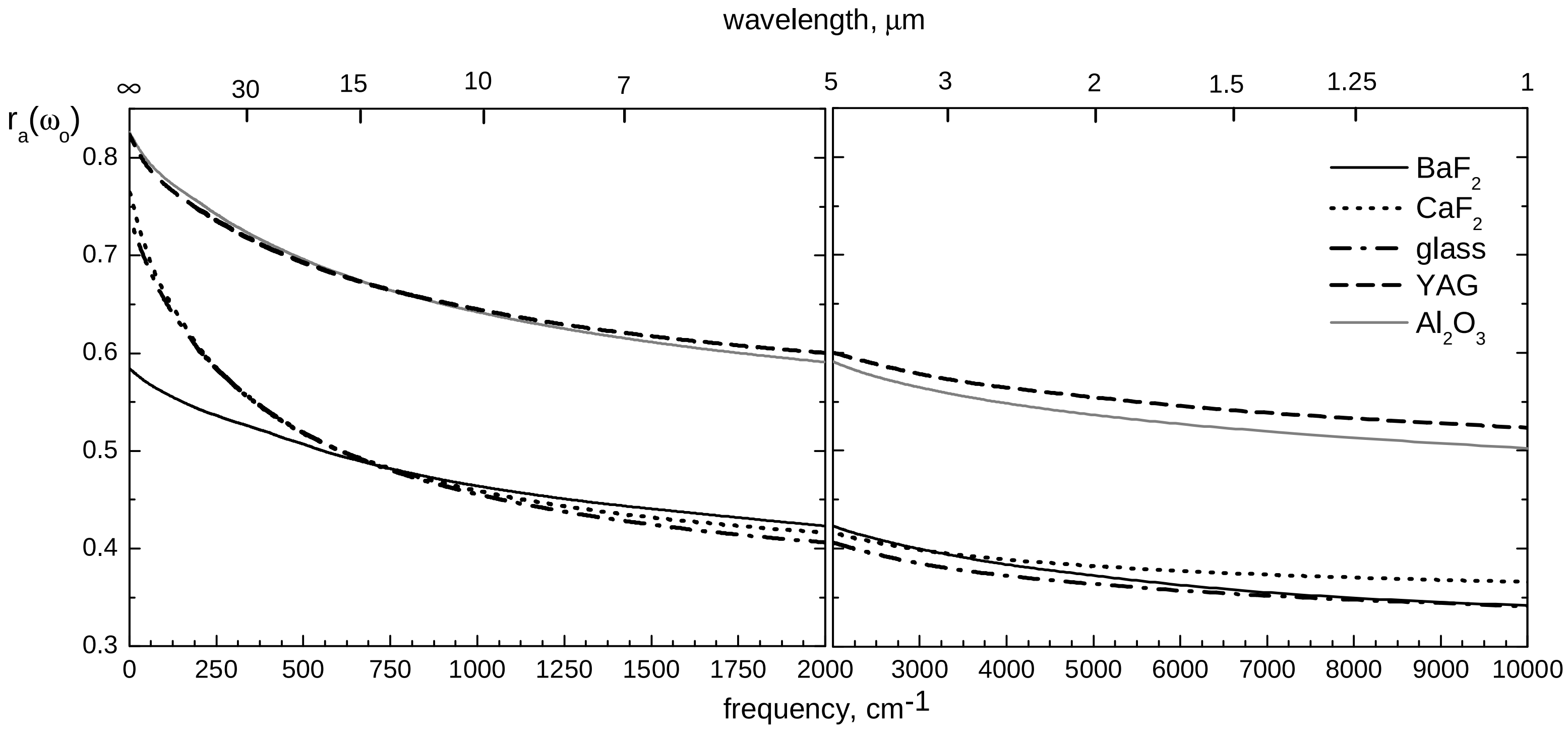

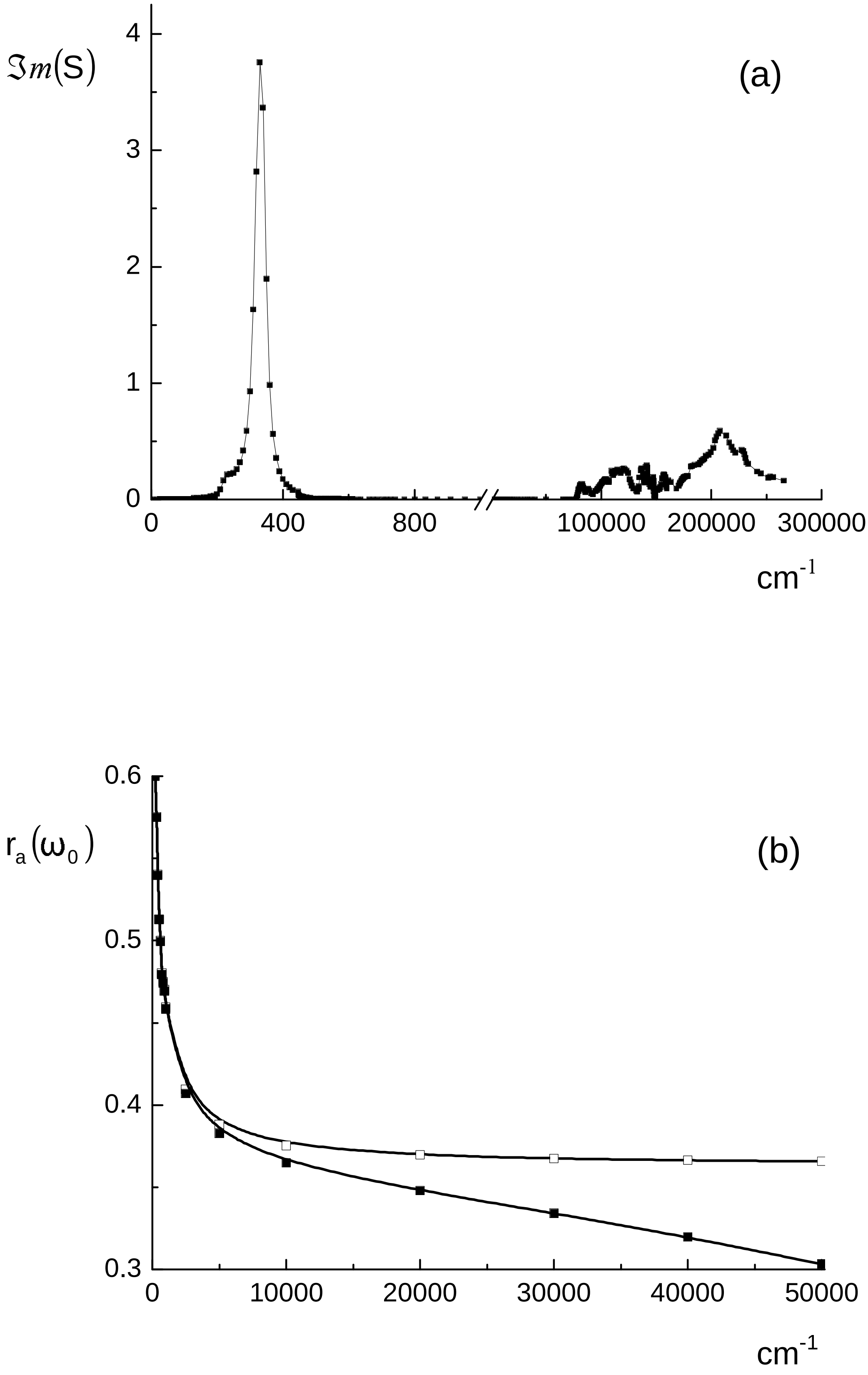